\documentclass[useAMS,usenatbib,usegraphicx]{mn2e}
\usepackage{lscape}
\usepackage{multicol}
\usepackage{longtable}
\usepackage{pdfpages}

\title[Infall Motions in Massive Star-Forming Regions]{Infall Motions in Massive Star-Forming Regions: Results from Years 1 \& 2 of the MALT90 Survey}
\author[Yu-Xin~He]{Yu-Xin~He$^{1,2}$\thanks{E-mail:heyuxin@xao.ac.cn}, Jian-Jun~Zhou$^{1,3}$\thanks{Email: zhoujj@xao.ac.cn}, Jarken~Esimbek$^{1,3}$, Wei-Guang~Ji$^{1,3}$, \newauthor Gang~Wu$^{1,2,3}$,  Xin-Di~Tang$^{1,3}$, Ye~Yuan$^{1,3}$, Da-Lei~Li$^{1,2}$ and W.A.~Baan$^{4,5}$\\
$^{1}$Xinjiang Astronomical Observatory, CAS, 150, Science 1-street, Urumqi, Xinjiang 830011, P. R. China\\
$^{2}$University of Chinese Academy of Sciences, 19A Yuquan Road, Beijing 100049, P. R. China\\
$^{3}$Key Laboratory of Radio Astronomy, Chinese Academy of Sciences, Nanjing, JiangSu 210008, China\\
$^{4}$Shanghai Astronomical Observatory, Chinese Academy of Sciences, Shanghai, Xuhui 203000, China\\
$^{5}$Netherlands Institute for Radio Astronomy, 7991 PD Dwingeloo, The Netherlands}

\begin{document}

\pagerange{\pageref{firstpage}--\pageref{lastpage}} \pubyear{2002}

\maketitle

\label{firstpage}

\begin{abstract}
Massive star-forming regions with observed infall motions are good sites for studying the birth of massive stars. In this paper, 405 compact sources have been extracted from the APEX Telescope Large Area Survey of the Galaxy (ATLASGAL) compact sources that also have been observed in the Millimetre Astronomy Legacy Team 90 GHz (MALT90) survey during Years 1 and 2. These observations are complemented with Spitzer GLIMPSE/MIPSGAL mid-IR survey data to help classify the elected star-forming clumps into three evolutionary stages: pre-stellar, proto-stellar and UCHII regions. The results suggest that 0.05 g cm$^{-2}$ is a reliable empirical lower bound for the clump surface densities required for massive-star formation to occur. The optically thick HCO$^{+}$(1-0) and HNC(1-0) lines, as well as the optically thin N$_{2}$H$^{+}$(1-0) line were used to search for infall motions toward these sources. By analyzing the asymmetries of the optically thick HCO$^{+}$(1-0) and HNC(1-0) lines and the mapping observations of HCO$^{+}$(1-0), a total of 131 reliable infall candidates have been identified. The HCO$^{+}$(1-0) line shows the highest occurrence of obvious asymmetric features, suggesting that it may be a better infall motion tracer than other lines such as HNC(1-0). The detection rates of infall candidates toward pre-stellar, proto-stellar and UCHII clumps are 0.3452, 0.3861 and 0.2152, respectively. The relatively high detection rate of infall candidates toward UCHII clumps indicates that many UCHII regions are still accreting matter. The peak column densities and masses of the infall candidates, in general, display a increasing trend with progressing evolutionary stages. However, the rough estimates of the mass infall rate show no obvious variation with evolutionary stage.
\end{abstract}


\section{Introduction}
Gravitational infall is a key part of the massive-star formation process, contributing via competitive accretion \citep{2004MNRAS.349..735B} and core accretion \citep{2003ApJ...585..850M}. These processes occur in high-mass young stellar objects (YSOs) at early evolutionary stages and continue their presence until later stage UCHII regions \citep{2003ApJ...599.1196K,2005ApJ...630..987S}. As shown by theoretical works \citep[e.g.][]{1996ApJ...462..874J,2002ApJ...569..846Y,2009ApJ...699..230G}, infall motion is critical for initiating high-mass star formation, while also maintaining accretion flows to feed the stellar mass during subsequent evolutionary stages. Simulations, theory and observations are converging to the idea that the collapse and outflow phenomenon is universal, covering the full range of stellar mass scales from brown dwarfs to massive stars. Moreover, infall and outflow motions should be closely related and interact with each other throughout the star formation history \citep{2014arXiv1401.2219L}. Thus infall candidates can serve as good sources in which to study massive star formation and gas dynamics in molecular cores.

Infall-motion surveys of high-mass star-forming regions have been reported in several recent papers \citep{2007ApJ...669L..37W,2007ApJ...663.1092K,2009MNRAS.392..170S,2011ApJ...740...40R,2012A&A...538A.140K,2013A&A...549A...5R}. However, high-mass infall candidates are still few. Further observations are needed to better constrain the physical properties of the infall, including its spatial distribution, mass infall rate, chemical effect, and understanding its relation with respect to other dynamical processes, including outflow, disk accretion and core fragmentation \citep{2012MNRAS.422.1098R}.

In this paper, 405 compact sources have been selected based on the results of the APEX Telescope Large Area Survey of the Galaxy (ATLASGAL) survey  \citep{2009A&A...504..415S} and the Millimetre Astronomy Legacy Team 90 GHz (MALT90) survey \citep{2013PASA...30...57J}. Infall motions toward these sources will be identified and studied using two optically thick lines HCO$^{+}$(1-0) and HNC(1-0), and one optically thin line, N$_{2}$H$^{+}$(1-0). The mapping observations of these surveys allow the identification of the most probable infall candidates. A brief introduction for the ATLASGAL survey and the sample selection are presented in Section 2.1, the MALT90 survey and Spitzer survey in Section 2.2 and 2.3, and the classification of the sample sources in Section 2.4. In Section 3 their kinematic distances, peak column densities and masses are calculated, and in Section 4 the infall candidates are identified and investigated. The conclusions are summarized in Section 5.

\section[]{DATA}
\subsection{The ATLASGAL survey}

The ATLASGAL is the first systematic survey of the inner Galactic plane at sub-millimeter wavelengths. Using the 12 m APEX telescope, the aim of this survey was to study continuum emission from the highest density regions of dust at 345 GHz. The angular resolution of the APEX telescope at this frequency is 19$^{\prime\prime}$.2. The typical pointing r.m.s. error was measured to be $\sim$4$^{\prime\prime}$, and the r.m.s. of the images is 50-70 mJy beam$^{-1}$ \citep{2009A&A...504..415S}.

\citet{2013A&A...549A..45C} published one catalog containing 6639 compact sources located in the range of $330^{\circ}$ $\leq$ $l$ $\leq$ $21^{\circ}$ and $|b|$ $\leq$ $1.5^{\circ}$ (see following). The data from this catalog has been complemented with observations from the MALT90 survey (years 1 and 2) \citep{2013PASA...30...57J}. The source sample has been selected using the following criteria: (i) detected N$_{2}$H$^{+}$(1-0), HNC(1-0), and HCO$^{+}$(1-0) emissions with signal-to-noise ratio larger than 3; (ii) a peak flux at 870 $\mu$m above 6$\sigma$, which corresponds to a flux density of $\sim$ 0.4 Jy beam$^{-1}$; and (iii) an angular distance between any two sources larger than the Mopra beam size (36$^{\prime\prime}$ at 90 GHz). These criteria ensure that the sources are not contaminated by emission from an adjacent clump. Using these criteria, 405 compact sources with good molecular lines have been selected for a sample.

\subsection{The MALT90 survey}
 The MALT90 Survey is a large international project that obtains molecular line maps in order to characterize the physical and chemical conditions of high-mass star formation regions over a wide range of evolutionary stages. The sample for this survey is a sub-sample of the ATLASGAL catalog. The angular resolution of the 22 m Mopra radio telescope at 90 GHz is 36$^{\prime\prime}$ \citep{2013PASA...30...57J}. The MALT90 data has been obtained from the online archive\footnote{http://atoa.atnf.csiro.au/MALT90/}. The data were reduced by the software GILDAS (Grenoble Image and Line Data Analysis Software).

\subsection{The Spitzer surveys}
The Galactic Legacy Infrared Mid-Plane Survey Extraordinaire (GLIMPSE) survey is a Spitzer/IRAC Legacy survey of the Galactic mid-plane \citep{2003PASP..115..953B,2009PASP..121..213C} at 3.6, 4.5, 5.8, and 8.0 $\mu$m, respectively. The angular resolution is better than 2$^{\prime\prime}$ at all wavelengths. The MIPS/Spitzer Survey of the Galactic Plane (MIPSGAL) is a survey of the Galactic plane at 24 and 70 $\mu$m using the Multiband Imaging Photometer aboard the Spitzer Space Telescope (MIPS). The angular resolution at 24 and 70 $\mu$m is 6$^{\prime\prime}$ and 18$^{\prime\prime}$ \citep{2009PASP..121...76C}. The highly reliable point source catalogs released from the GLIMPSE survey and the mosaicked images of MIPSGAL at 24 $\mu$m have been used in the following analysis.

\subsection{Classification}
The clumps, which contain objects (in projection) obeying the criteria (a point-source should have [4.5] - [5.8] $>$ 1.0 and be detected at 8 $\mu$m) or (a point-source should have [4.5] - [5.8] $>$ 0.7, [3.6] - [4.5] $>$ 0.7 and be detected at 8 $\mu$m), were assumed to host star-forming activities \citep{2008ApJ...674..336G}. Any clumps associated with 24 $\mu$m point sources were also assumed to host a forming star. As a MIPSGAL 24 $\mu$m point source catalog has not been published, the STARFINDER algorithm \citep{2000A&AS..147..335D} has been used to search for point sources in the 24 $\mu$m MIPSGAL images. Sources have been extracted that had an S/N ratio better than 7, resulting in 280 clumps that were associated with star formation. Of these 280 star-forming clumps, 78 are associated with IRAS sources, and 64 out of these 78 clumps were IRAS sources that satisfied the criteria for being an UCHII region \citep{1989ApJS...69..831W}. Fifteen of the 280 star-forming clumps were identified as UCHII regions in the literature, (see the last column of Table 1 for their corresponding references). The remaining 201 star-forming clumps were classified as proto-stellar clumps. Eighty-four clumps having no star-formation properties were classified as pre-stellar clumps. Column 10 of Table 1 lists the evolutionary stages of all clumps in our sample. It should be noted that no attempts have been made to determine the evolutionary stage of 41 clumps associated with saturated 24 $\mu$m sources or photo-dissociation regions; these clumps are indicated as "Non" in Table 1.

\begin{table*}
\scriptsize
 \centering
  \begin{minipage}{175mm}
   \caption{Examples of the derived clump parameters. The columns are as follows: (1) and (2) ATLASGAL and Clump names; (3) peak submillimetre emission; (4) integrated submillimetre emission; (5) heliocentric distance; (6) references and distance; (7) effective physical radius; (8) column density; (9) clump mass derived from the integrated 870$\,\umu$m emission; (10) Spitzer classification; and (11) references and classification.   The full table is available online.}
    \begin{tabular}{rrrrrrrrrrrrrrrrrrrrrrrr}
      \hline
      \multicolumn{1}{c}{ATLASGAL} & \multicolumn{1}{c}{Clump$^{\rm{a}}$} & \multicolumn{1}{c}{Peak flux} & \multicolumn{1}{c}{Int. flux} & \multicolumn{2} {c} {Distance} & \multicolumn{1}{c}{Radius} & \multicolumn{1}{c}{Log($N(H_2\,)$)} & \multicolumn{1}{c}{Log($M_{clump}\,$)} & \multicolumn{2}{c}{$Spitzer$ $classification^{\rm{b}}$}\\
      \multicolumn{1}{c}{name} & \multicolumn{1}{c}{name} & \multicolumn{1}{c}{($Jy\ beam^{-1}\,$)} & \multicolumn{1}{c}{($Jy$)} & \multicolumn{1}{c}{($kpc$)} & \multicolumn{1}{c}{Ref.} & \multicolumn{1}{c}{($pc$)} & \multicolumn{1}{c}{($cm^{-2}\,$)} & \multicolumn{1}{c}{($M_{\odot}$)\,} & & \multicolumn{1}{c}{Ref.}\\
      \multicolumn{1}{c}{(1)} & \multicolumn{1}{c}{(2)} & \multicolumn{1}{c}{(3)} & \multicolumn{1}{c}{(4)} & \multicolumn{1}{c}{(5)} & \multicolumn{1}{c}{(6)} & \multicolumn{1}{c}{(7)} & \multicolumn{1}{c}{(8)} & \multicolumn{1}{c}{(9)} & \multicolumn{1}{c}{(10)} & \multicolumn{1}{c}{(11)}\\
      \hline
      AGAL003.416$-$00.354  &  *G003.416$-$0.354   &    0.83    &    10.4     &   $-$     &   $-$   &  $-$    &  22.56   &    $-$     &  Prestellar  & $-$  \\
AGAL003.434$-$00.572  &  *G003.434$-$0.572   &    0.74    &    3.58     &   $-$     &   $-$   &  $-$    &  22.51   &    $-$     &  Prestellar  & $-$  \\
AGAL005.917$-$00.311  &  *G005.917$-$0.311   &    0.56    &    4.54     &   16.12   &   3     &  1.17   &  22.39   &    4.07    &  Prestellar  & $-$  \\
AGAL007.234$-$00.101  &  *G007.234$-$0.101   &    1.02    &    15.06    &   3.35    &   3     &  0.75   &  22.65   &    3.22    &  Prestellar  & $-$  \\
AGAL010.288$-$00.124  &  *G010.288$-$0.124   &    4.4     &    10.74    &   3.55    &   2     &  0.26   &  23.29   &    3.13    &  Prestellar  & $-$  \\
AGAL012.776$-$00.211  &  *G012.776$-$0.211   &    1.93    &    14.73    &   3.57    &   3     &  0.59   &  22.93   &    3.27    &  Prestellar  & $-$  \\
AGAL013.169+00.077    &  *G013.169+0.077     &    1.01    &    5.14     &   4.27    &   3     &  0.39   &  22.65   &    2.97    &  Prestellar  & $-$  \\
AGAL013.276$-$00.334  &  *G013.276$-$0.334   &    1       &    16.03    &   3.83    &   3     &  1.21   &  22.64   &    3.37    &  Prestellar  & $-$  \\
AGAL014.626$-$00.562  &  *G014.626$-$0.562   &    1.18    &    23.24    &   2.1     &   3     &  1.00   &  22.72   &    3.00    &  Prestellar  & $-$  \\
AGAL331.029$-$00.431  &  *G331.029$-$0.431   &    1.98    &    26.23    &   3.92    &   3     &  1.03   &  22.94   &    3.60    &  Prestellar  & $-$  \\
      \hline
     \end{tabular}
     \medskip
     $^{\rm{a}}$ Sources are named by galactic coordinates of ATLASGAL sources:  An $\ast$ indicates infall candidates.\\
     $^{\rm{b}}$ A $\dagger$ indicates source with IRAS counterpart.\\
  References --- Distance: (1) \citet{2013MNRAS.431.1752U}, (2) \citet{2013MNRAS.435..400U}, (3) this paper, (4) \citet{2005A&A...429..945M}, (5) \citet{2013A&A...550A..21S}, (6) \citet{2004A&A...426...97F}, (7) \citet{2012ApJS..202....1H}, (8) \citet{2012A&A...547A..49R}, (9) \citet{2003A&A...397..133R}, (10) \citet{2006MNRAS.366.1096B}, (11) \citet{1998A&AS..132..211H}, (12) \citet{1982ApJS...49..183B}.\\
  References --- Spitzer classification: (1) \citet{1994ApJS...91..347B}, (2) \citet{2000ApJ...530..371F}, (3) \citet{1997MNRAS.291..261W}, (4) \citet{2011ApJS..194...32A}, (5) \citet{2001ApJ...549..979K}, (6) \citet{1987A&A...181..378C}, (7) \citet{2009A&A...501..539U}, (8) \citet{2007ApJ...669L..37W}\\

  \end{minipage}
\end{table*}
\normalsize

\section{physical properties of the sample}
\subsection{Kinematic distances}
Distances of 132 clumps have been obtained from the literature. For the other 273 clumps, the Galactic rotation model of \citet{2009ApJ...700..137R} and the radial velocities of N$_{2}$H$^{+}$(1-0) were used to estimate their kinematic distances. It should be noted that, if one source is located outside the solar circle (i.e. $>$ 8.5 kpc from the Galactic Centre) or is at a tangential point, we will calculate one unique distance. However, if one source is located within the solar circle (i.e. $<$ 8.5 kpc from the Galactic Centre), two possible distances are obtained (one near, one far). This degeneracy is commonly referred to as the kinematic distance ambiguity (KDA). Here the HI self-absorption technique \citep{2006MNRAS.366.1096B} is used to resolve the KDA. For this purpose HI spectra have been extracted from the Southern Galactic Plane Survey \citep[SGPS:][]{2005ApJS..158..178M} and the VLA Galactic Plane Survey \citep[VGPS:][]{2006AJ....132.1158S}. If a clump is at the near distance, cold and dense HI contained therein will absorb warmer HI background line emission, and the spectrum of HI will show self-absorption, whereas any clumps at the farther distance will not display any self-absorption as there is not any background radiation to absorb \citep{2006MNRAS.366.1096B}. Using this method, we determined the kinematic distance to 257 clumps (see Table 1). For the remaining 16 clumps, 14 of them are located within the solar circle and do not have any HI data, so we are unable to resolve their KDAs. For another two clumps, G331.374$-$0.314 and G353.019+0.976, although their observed N$_{2}$H$^{+}$(1-0) peak lies in a HI trough, HI spikes are present within the trough making it impossible to resolve their KDAs.  As an example of this procedure, Fig. 1 shows an unambiguous near-distance solution using the N$_{2}$H$^{+}$(1-0) emission line that coincides with a HI absorption (right panel), and a far-distance clump with no HI absorption located at the N$_{2}$H$^{+}$(1-0) peak velocity (left panel). The N$_{2}$H$^{+}$(1-0) and overlaid HI spectral profiles of all 273 clumps are available online.

\begin{figure}
 \centering
  \includegraphics[width=0.45\textwidth]{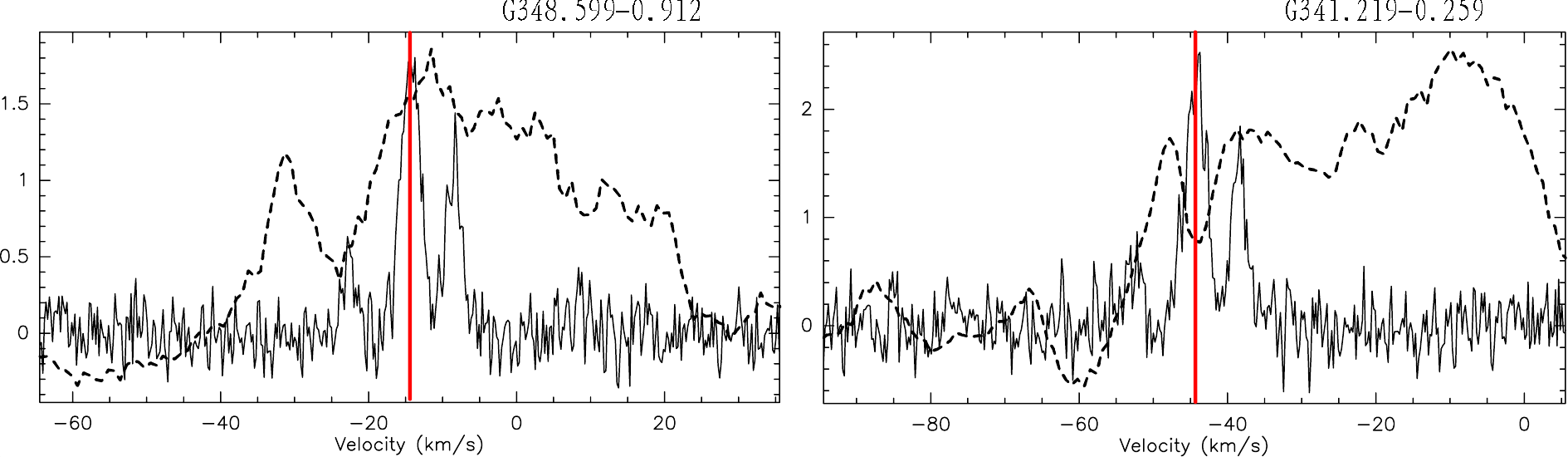}
   \caption{Representative spectra for the near and far distances obtained using the HI self-absorption technique. The solid line represents the N$_{2}$H$^{+}$(1-0) spectra overlaid with the HI 21-cm data (bold dashed line), with the HI data scaled to the peak of N$_{2}$H$^{+}$(1-0). The vertical solid red line indicates the velocity of the clump. Left panel represents a far-distance object with no HI self-absorption at the N$_{2}$H$^{+}$(1-0) peak velocity, while right panel shows a clear near-distance solution with the N$_{2}$H$^{+}$(1-0) emission line coinciding with a HI absorption. }
\end{figure}

\subsection{Peak column density}
The peak flux density at 870 $\mu$m of each clump (Table 1) was used to derive the beam-averaged H$_{2}$ column density via the formula $N_{H_2}={{S_\nu}R\over{B_\nu}(T_D){\Omega}{\kappa_\nu}{\mu}{m_H}}$, where S$_{\nu}$ is the peak 870 $\mu$m flux density, R is the gas-to-dust mass ratio (assumed to be 100), $\Omega$ is the beam solid angle, $\mu$ is the mean molecular weight of the interstellar medium assumed to be equal to 2.8, m$_{H}$ is the mass of a hydrogen atom, B$_{\nu}$ is the Planck function for dust temperature T$_{D}$, and $\kappa$$_{\nu}$ is the dust-absorption coefficient taken as 1.85 cm$^{2}$ g$^{-1}$ \citep[interpolated to 870 $\mu$m from Col. 5 of Table 1, ][]{1994A&A...291..943O}. Following the results of  \citet{2013ApJ...777..157H}, the temperatures of pre-stellar, proto-stellar and UCHII clumps are assumed to be 13.9 K, 17.9 K, and 26 K, respectively.

The distributions of the 870 $\mu$m peak flux densities for the pre-stellar, proto-stellar and UCHII clumps are shown in Fig. 2 (a), which have median values of 1.13, 1.98 and 4.59 Jy beam$^{-1}$, respectively. This may be ascribed to increasing dust temperature towards the center of each clump, where star-formation processes heat the dust. Fig. 2 (b) and (c) present the H$_{2}$ column density distributions of the pre-stellar, proto-stellar and UCHII region clumps, and the complete sample, respectively. The median values of these distributions are indicated by the dashed black vertical line, which are 5.01$\times$10$^{22}$ cm$^{-2}$, 5.75$\times$10$^{22}$ cm$^{-2}$, and 7.94$\times$10$^{22}$ cm$^{-2}$, respectively. The H$_{2}$ column densities display a trend of increasing with evolutionary stage. The mean, median, and standard deviation of each distribution are summarized in Table 5.

\begin{figure}
 \centering
  \includegraphics[width=0.45\textwidth]{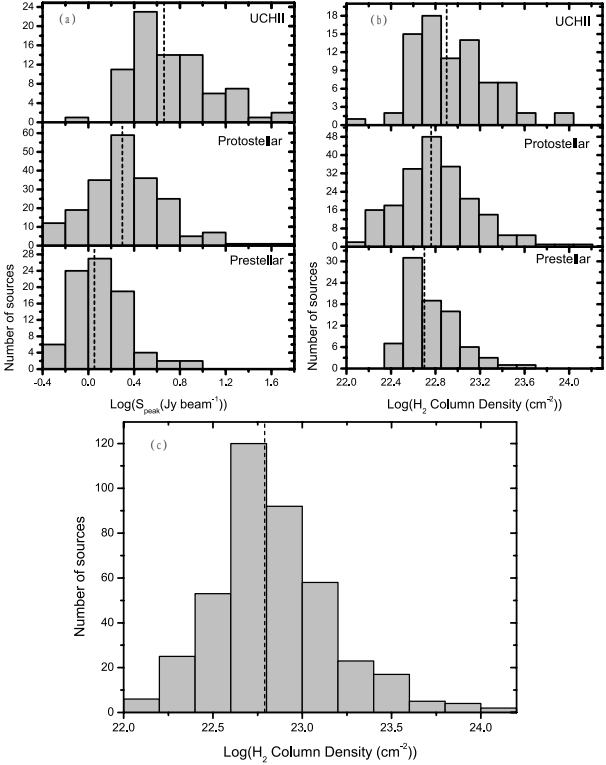}
   \caption{The 870 $\mu$m emission properties. (a): The distributions of 870 $\mu$m peak flux densities for all classified sources separated by evolutionary stage.  (b): The distributions of beam-averaged H$_{2}$ column densities derived from 870 $\mu$m dust emission for all classified sources separated by evolutionary stage, where the vertical dashed black line indicates the median value for each stage. (c): The distribution of beam-averaged H$_{2}$ column densities for the whole sample. Median value is indicated by the dashed black vertical lines.}
\end{figure}

\begin{figure}
 \centering
  \includegraphics[width=0.45\textwidth]{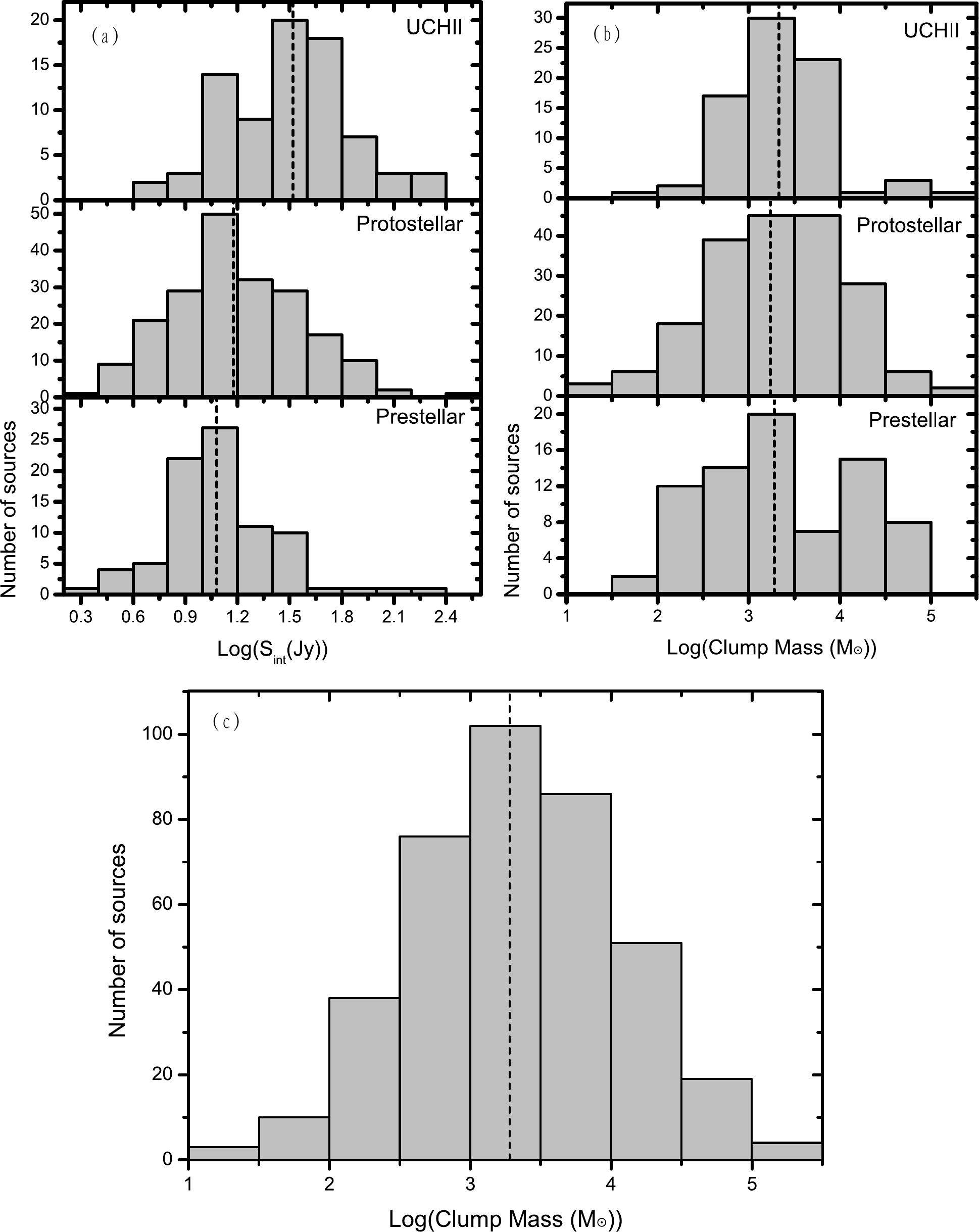}
   \caption{The 870 $\mu$m dust mass determinations. (a): 870 $\mu$m integrated flux densities for all classified sources separated by evolutionary stage.  (b): The dust mass distribution for all classified sources separated by evolutionary stage.  (c): The dust mass distribution for the whole sample. Median values are indicated by the dashed black vertical lines.}
\end{figure}

\subsection{Clump mass}
Assuming that the dust emission is optically thin in the sub-millimeter continuum, masses of clumps with known distances can be calculated from the dust continuum emission via $M={{D^2}{S_\nu}{R}\over{B_\nu}(T_D){\kappa_\nu}}$, where S$_{\nu}$ is the integrated 870 $\mu$m flux and D is the heliocentric distance to the source, R, B$_{\nu}$, T$_{D}$, and $\kappa$$_{\nu}$ are the same as in section 3.2. Fig. 3 (a) shows the distributions of the pre-stellar, proto-stellar and UCHII clumps as a function of the integrated 870 $\mu$m flux density. Their corresponding median values are 12.3, 14.98 and 33.06 Jy, respectively. The median values of the mass distributions of the pre-stellar, proto-stellar and UCHII clumps are 1905.5, 1737.8 and 2138.0 M$_{\odot}$, respectively (Fig. 3 (b)). The masses of the pre-stellar, proto-stellar and UCHII clumps show similar distributions. The statistical parameters of these distributions are summarized in Table 5. More than 96.6\% of all clumps have masses larger than 100 M$_{\odot}$, and the median value is 1905.5 M$_{\odot}$ (Fig. 3 (c)).

\citet{2010ApJ...716..433K} found that high-mass star-forming regions obey the mass-size relationship m(r)$\geq$580M$_{\odot}$(R$_{eff}$ pc$^{-1}$)$^{1.33}$, which was confirmed by \citet{2013MNRAS.431.1752U}. Hence the relation may provide a suitable description for massive star formation. In the sample, 375 (96.4\%) clumps have masses larger than the limiting masses for their size, and they are promising sites of high-mass star formation. Fig. 4 displays the mass-size relationship for the 389 clumps, all of them are spatially resolved by the APEX beam. The dashed red line indicates the least-squares fit to 389 clumps expressed as empirical relation of Log(M$_{clump}$)=3.41$\pm$0.01+(1.78$\pm$0.03)$\times$Log(R$_{eff}$) with correlation coefficient of 0.94. The upper and lower red diagonal lines indicate constant surface densities, $\Sigma$(gas), of 1 g cm$^{-2}$ and 0.05 g cm$^{-2}$, respectively. All sources are located in the region $\Sigma$(gas) $>$ 0.05 g cm$^{-2}$. This is consistent with the result of \citet{2013MNRAS.431.1752U}. Therefore, high-mass star formation may take place when the surface density is larger than 0.05 g cm$^{-2}$.

\begin{figure}
 \centering
  \includegraphics[width=0.45\textwidth]{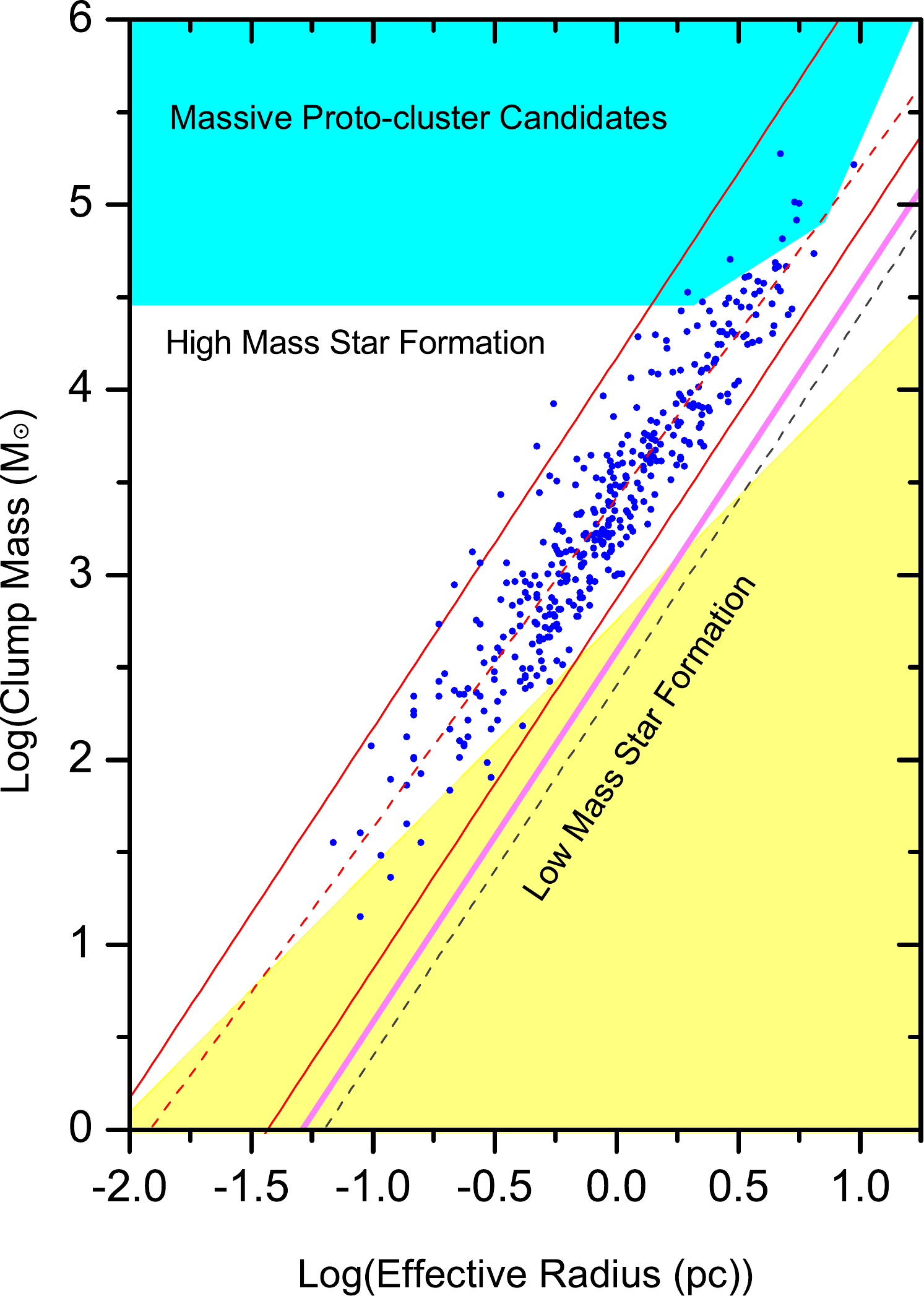}
   \caption{The mass-size relationship of 389 clumps that have determined mass values and are spatially resolved by the APEX beam (blue dots). The yellow shaded region shows the part of the parameter space found to be devoid of massive star formation that satisfies the relationship m(r)$\leq$580M$_{\odot}$(R$_{eff}$ pc$^{-1}$)$^{1.33}$) \citep[cf.][]{2010ApJ...723L...7K}. The green shaded region indicates the region in which young massive cluster progenitors are expected to be found \citep[i.e.][]{2012ApJ...758L..28B}. The dashed red line shows the result of linear least-squares fits to the 389 clumps with calculated mass values. The grey dashed line shows the sensitivity of the ATLASGAL survey, the upper and lower solid red line shows the surface densities of 1 and 0.05 g cm$^{-2}$.}
\end{figure}

Applying the criteria of \citet{2012ApJ...758L..28B} to the current sample results in seven young massive protocluster (MPC) candidates in the green-shaded area of Fig.4. Of these G345.504+0.347 and G351.774-0.537 have already been identified by \citet{2010ApJ...712.1137K}, while G008.691-0.401, G333.299-0.351, G338.459+0.024, G348.183+0.482 and G348.759-0.946 are newly identified in this work.

\section{Infall candidates}
\subsection{Identification of infall candidates}
The HCO$^{+}$(1-0), HNC(1-0) and N$_{2}$H$^{+}$(1-0) line profiles have been extracted at the centers of all 405 clumps as determined by the peak 870 $\mu$m emission of each clump. The line parameters were derived by Gaussian fitting. It should be noted that some clumps show asymmetric profiles that cannot be fitted by a single-Gaussian function directly. However, the upper portion of the highest peak of HCO$^{+}$(1-0) and HNC(1-0) profile shows a symmetric profile, which can be fitted by a single-Gaussian function. The uncertainties of the velocities at these peaks were estimated by the Gaussian fit. The optically thin N$_{2}$H$^{+}$(1-0) lines were fitted using the hyperfine fitting routines in $CLASS$ to determine their peak velocity and FWHM. The derived line parameters of all sources are given in Table 2.

\begin{table*}
\scriptsize
 \centering
  \begin{minipage}{155mm}
   \caption{Examples of the derived line parameters and profiles of the observed sources. Quantities in parentheses give the uncertainties in units of 0.01. The columns are as follows: (1) Clump names; (2) peak velocity of $HCO^{+}(1-0)$; (3) peak velocity of $HNC(1-0)$; (4) peak velocity of $N_{2}H^{+}(1-0)$; (5) FWHM of $N_{2}H^{+}(1-0)$; (6) asymmetry of $HCO^{+}(1-0)$; (7) asymmetry of $HNC(1-0)$; (8) profile of $HCO^{+}(1-0)$ and $HNC(1-0)$. The full table is available online. }
    \begin{tabular}{rrrrrrrrrrrrrrrrrrrrrrrr}
        \hline
        \multicolumn{1}{c}{Clump$^{\rm{a}}$} & \multicolumn{1}{c}{$V_{thick}$\,} & \multicolumn{1}{c}{$V_{thick}$\,} & \multicolumn{1}{c}{$V_{thin}$\,} & \multicolumn{1} {c} {$\Delta V$\,} & \multicolumn{1}{c}{$\delta v$\,} & \multicolumn{1}{c}{$\delta v$\,} & \multicolumn{1}{c}{Profile} \\
        \multicolumn{1}{c}{name} & \multicolumn{1}{c}{$HCO^{+}(1-0)$\,} & \multicolumn{1}{c}{$HNC(1-0)$\,)} & \multicolumn{1}{c}{$N_{2}H^{+}(1-0)$\,} & \multicolumn{1}{c}{$N_{2}H^{+}(1-0)$\,} & \multicolumn{1}{c}{$HCO^{+}(1-0)$\,} & \multicolumn{1}{c}{$HNC(1-0)$\,} & \\
        & \multicolumn{1}{c}{$km\ s^{-1}$\,} & \multicolumn{1}{c}{$km\ s^{-1}$\,} & \multicolumn{1}{c}{$km\ s^{-1}$\,} & \multicolumn{1}{c}{$km\ s^{-1}$\,} & & & \\
        \multicolumn{1}{c}{(1)} & \multicolumn{1}{c}{(2)} & \multicolumn{1}{c}{(3)} & \multicolumn{1}{c}{(4)} & \multicolumn{1}{c}{(5)} & \multicolumn{1}{c}{(6)} & \multicolumn{1}{c}{(7)}  & \multicolumn{1}{c}{(8)}\\
        \hline
      $*$G003.416$-$0.354 &  $-$27.17    (  24  )   &    $-$25.80    (  05  )   &  $-$25.28   (  32  )   &    3.31  (  46  )   &    $-$0.57  (  25  )   &  $-$0.16   (  13  )      &       B,N      \\
$*$G003.434$-$0.572 &  1.20        (  01  )   &    2.40        (  14  )   &  2.51       (  09  )   &    1.03  (  19  )   &    $-$1.28  (  33  )   &  $-$0.11   (  24  )      &       B,N      \\
$*$G005.917$-$0.311 &  9.32        (  02  )   &    9.67        (  03  )   &  10.09      (  11  )   &    1.51  (  19  )   &    $-$0.51  (  15  )   &  $-$0.28   (  13  )      &       B,B      \\
$*$G007.234$-$0.101 &  18.21       (  05  )   &    18.53       (  07  )   &  19.13      (  16  )   &    2.11  (  31  )   &    $-$0.44  (  16  )   &  $-$0.28   (  15  )      &       B,B      \\
$*$G010.288$-$0.124 &  13.05       (  11  )   &    12.69       (  12  )   &  13.86      (  04  )   &    2.74  (  08  )   &    $-$0.30  (  06  )   &  $-$0.43   (  07  )      &       B,B      \\
$*$G012.776$-$0.211 &  34.81       (  09  )   &    35.12       (  31  )   &  35.47      (  05  )   &    1.68  (  10  )   &    $-$0.39  (  11  )   &  $-$0.21   (  23  )      &       B,N      \\
$*$G013.169+0.077   &  48.27       (  03  )   &    49.73       (  21  )   &  49.72      (  08  )   &    2.35  (  16  )   &    $-$0.62  (  09  )   &  0.00      (  12  )      &       B,N      \\
$*$G013.276$-$0.334 &  39.94       (  24  )   &    40.26       (  05  )   &  41.26      (  06  )   &    1.32  (  26  )   &    $-$1.00  (  43  )   &  $-$0.76   (  23  )      &       B,B      \\
$*$G014.626$-$0.562 &  16.71       (  02  )   &    17.00       (  09  )   &  17.96      (  04  )   &    1.90  (  09  )   &    $-$0.66  (  06  )   &  $-$0.51   (  09  )      &       B,B      \\
$*$G331.029$-$0.431 &  $-$65.82    (  11  )   &    $-$65.36    (  13  )   &  $-$64.37   (  07  )   &    3.01  (  12  )   &    $-$0.48  (  08  )   &  $-$0.33   (  08  )      &       B,B      \\
      \hline
     \end{tabular}
     \medskip
      $^{\rm{a}}$ Sources are named by galactic coordinates of ATLASGAL sources:  An $\ast$ indicates infall candidates.\\
  NOTE. The HCO$^{+}$(1-0), and HNC(1-0) profiles are evaluated as follows: B denotes a blue profile, R denotes a red profile, and N denotes neither blue nor red.\\
   \end{minipage}
\end{table*}
\normalsize

Following the method of \citet{1997ApJ...489..719M}, 150 blue profiles were identified using the HCO$^{+}$(1-0) lines, 99 blue profiles were identified using the HNC(1-0) lines, and 87 sources show blue profiles in both the HCO$^{+}$(1-0) and HNC(1-0) spectra. The profile asymmetries are given in Table 2. Any sources that display blue profiles are possible infall candidates. However, rotation might cause a blue asymmetric line profile at one side of the rotation axis, but at the same time the red asymmetry could appear at the other side of the rotation axis. Outflow could also cause blue and red asymmetry on opposite directions. Checking the spatial variation of profile asymmetries of these sources can help us to identify reliable infall candidates.

The source G012.418+0.506 is a typical example of an infall candidate (Fig. 5). The right panel of Fig.5 shows N$_{2}$H$^{+}$(1-0), HNC(1-0) and HCO$^{+}$(1-0) spectra detected toward the peak position of this clump. An obvious blue asymmetry is present in both of the HCO$^{+}$(1-0) and HNC(1-0) spectra. It is seen that the HCO$^{+}$(1-0) spectra displays a blue asymmetry in all of the mapping regions (left panel of Fig. 5). Conversely, the source G337.406-0.402 appears to be a typical example of a blue asymmetry caused by rotation (Fig. 6). HCO$^{+}$(1-0) spectra detected toward the peak position of this clump display a blue asymmetry (right panel of Fig. 6). However, the profile asymmetry of HCO$^{+}$(1-0) shows spatial variation across the mapping region. Because of the rotation, the extended blue asymmetry of the HCO$^{+}$(1-0) profiles reverses to an extended red asymmetry with respect to the northwest-southeast axis through the center of the clump (left panel of Fig. 6). The figures of the other infall candidates are available online.

\begin{figure}
 \centering
  \includegraphics[width=0.45\textwidth]{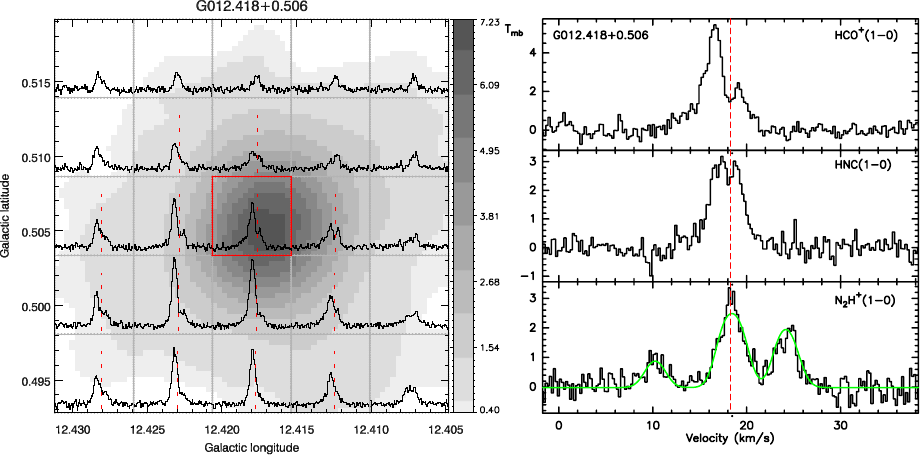}
   \caption{Example of an infall source G012.418+0.506. The left panel: the HCO$^{+}$(1-0) map grid (gridded to 1/2 beam size) superposed on the 870 $\mu$m continuum emission map (starting from a flux density of 0.4 Jy beam$^{-1}$, which corresponds to a peak flux above 6$\sigma$). The right panel: the extracted spectra of HCO$^{+}$(1-0), HNC(1-0) and N$_{2}$H$^{+}$(1-0) from the central position of this clump (red square on the left panel). The dashed red lines on the profiles indicate the velocities of N$_{2}$H$^{+}$(1-0).}
\end{figure}

\begin{figure}
 \centering
  \includegraphics[width=0.45\textwidth]{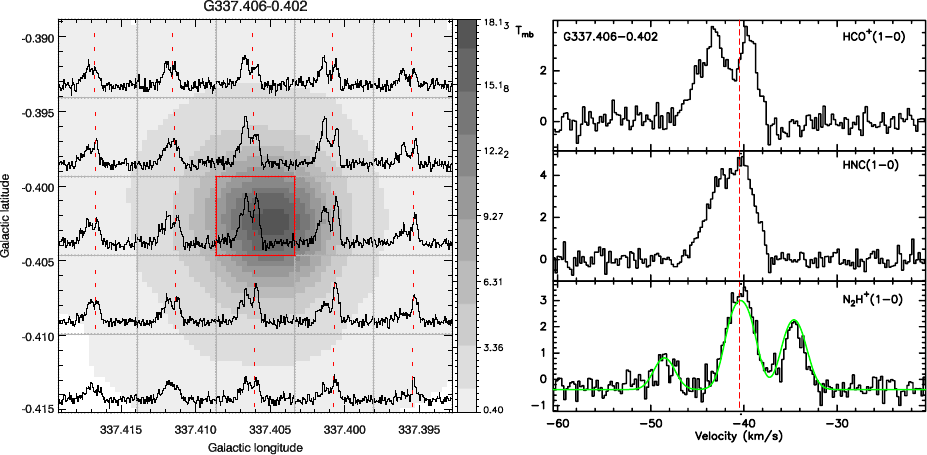}
   \caption{Example of the rotation source G337.406-0.402. Legend is the same as for Figure 5.}
\end{figure}

Finally, a total of 29 infall candidates were identified among the 84 pre-stellar clumps; 78 infall candidates among the 201 proto-stellar clumps; 17 infall candidates among the 79 UCHII region clumps; and seven infall candidates among the 41 clumps that do not have a classification (Table 1). The parameters of all 131 reliable infall candidates are listed in Table 3. The detection rates of the infall candidates are 0.3452, 0.3881 and 0.2152 for pre-stellar, proto-stellar, and UCHII clumps, respectively (Table 4).

\begin{table*}
\scriptsize
 \centering
  \begin{minipage}{125mm}
   \caption{Mass infall rates of infall candidates.}
    \begin{tabular}{lrrlrr}
      \hline
      \multicolumn{1}{c}{Clump}              &  \multicolumn{1}{c}{$\dot{M}$}                         & \multicolumn{1}{c}{$Spitzer$}       &
       \multicolumn{1}{c}{Clump}             &  \multicolumn{1}{c}{$\dot{M}$}                         &  \multicolumn{1}{c}{$Spitzer$}      \\
      \multicolumn{1}{c}{name}               &  \multicolumn{1}{c}{0.01 M$_{\odot}$ yr$^{-1}$}        & \multicolumn{1}{c}{classification}  &
       \multicolumn{1}{c}{name}              &  \multicolumn{1}{c}{0.01 M$_{\odot}$ yr$^{-1}$}        & \multicolumn{1}{c}{classification}  \\
      \hline
      G003.416$-$0.354   &  $-$    &    Prestellar     &      G333.076$-$0.417 &  0.30    &   Protostellar  \\     
G003.434$-$0.572   &  $-$    &    Prestellar     &      G333.386+0.032   &  1.49    &   Protostellar  \\     
G005.917$-$0.311   &  2.35   &    Prestellar     &      G334.651+0.469   &  0.38    &   Protostellar  \\     
G007.234$-$0.101   &  0.62   &    Prestellar     &      G335.262$-$0.114 &  0.50    &   Protostellar  \\     
G010.288$-$0.124   &  1.30   &    Prestellar     &      G336.411$-$0.256 &  1.70    &   Protostellar  \\     
G012.776$-$0.211   &  0.64   &    Prestellar     &      G336.958$-$0.224 &  1.74    &   Protostellar  \\     
G013.169+0.077     &  1.05   &    Prestellar     &      G337.176$-$0.032 &  1.32    &   Protostellar  \\     
G013.276$-$0.333   &  0.78   &    Prestellar     &      G337.258$-$0.101 &  2.36    &   Protostellar  \\     
G014.626$-$0.562   &  0.39   &    Prestellar     &      G337.671$-$0.192 &  0.31    &   Protostellar  \\     
G331.029$-$0.431   &  1.72   &    Prestellar     &      G337.761$-$0.339 &  0.63    &   Protostellar  \\     
G331.051$-$0.419   &  1.01   &    Prestellar     &      G337.916$-$0.477 &  2.40    &   Protostellar  \\     
G331.374$-$0.314   &  $-$    &    Prestellar     &      G338.916+0.382   &  3.40    &   Protostellar  \\     
G331.496$-$0.079   &  2.68   &    Prestellar     &      G338.926+0.634   &  3.41    &   Protostellar  \\     
G332.254$-$0.056   &  0.52   &    Prestellar     &      G339.584$-$0.127 &  1.20    &   Protostellar  \\     
G333.449$-$0.182   &  0.91   &    Prestellar     &      G339.924$-$0.084 &  2.03    &   Protostellar  \\     
G333.524$-$0.269   &  1.30   &    Prestellar     &      G341.217$-$0.212 &  1.36    &   Protostellar  \\     
G337.927$-$0.432   &  0.57   &    Prestellar     &      G341.219$-$0.259 &  0.71    &   Protostellar  \\     
G338.384+0.011     &  0.78   &    Prestellar     &      G341.236$-$0.271 &  1.31    &   Protostellar  \\     
G340.206$-$0.049   &  1.30   &    Prestellar     &      G341.266$-$0.302 &  1.04    &   Protostellar  \\     
G340.261$-$0.196   &  0.94   &    Prestellar     &      G341.942$-$0.166 &  1.03    &   Protostellar  \\     
G341.932$-$0.174   &  1.03   &    Prestellar     &      G343.489$-$0.064 &  0.33    &   Protostellar  \\     
G342.114+0.504     &  0.34   &    Prestellar     &      G343.489$-$0.416 &  0.08    &   Protostellar  \\     
G350.772+0.796     &  0.09   &    Prestellar     &      G343.738$-$0.112 &  0.34    &   Protostellar  \\     
G350.784+0.801     &  0.07   &    Prestellar     &      G343.902$-$0.672 &  1.25    &   Protostellar  \\     
G351.173+0.632     &  0.43   &    Prestellar     &      G344.102$-$0.662 &  1.12    &   Protostellar  \\     
G351.434+0.676     &  0.91   &    Prestellar     &      G346.369$-$0.647 &  0.50    &   Protostellar  \\     
G352.511+0.776     &  0.09   &    Prestellar     &      G348.533$-$0.982 &  4.06    &   Protostellar  \\     
G354.629$-$0.611   &  0.12   &    Prestellar     &      G348.579$-$0.919 &  0.52    &   Protostellar  \\     
G355.196$-$0.486   &  0.20   &    Prestellar     &      G348.599$-$0.912 &  4.13    &   Protostellar  \\     
G000.484$-$0.702   &  $-$    &    Protostellar   &      G350.411$-$0.064 &  1.52    &   Protostellar  \\     
G004.627$-$0.666   &  1.26   &    Protostellar   &      G350.521$-$0.349 &  0.35    &   Protostellar  \\     
G005.359+0.014     &  0.12   &    Protostellar   &      G350.757+0.942   &  3.00    &   Protostellar  \\     
G005.617$-$0.082   &  0.90   &    Protostellar   &      G351.444+0.659   &  3.43    &   Protostellar  \\     
G005.834$-$0.514   &  0.48   &    Protostellar   &      G352.502+0.784   &  0.26    &   Protostellar  \\     
G006.216$-$0.609   &  0.74   &    Protostellar   &      G353.012+0.557   &  0.08    &   Protostellar  \\     
G007.166+0.131     &  1.45   &    Protostellar   &      G353.089+0.444   &  0.11    &   Protostellar  \\     
G008.411$-$0.347   &  0.83   &    Protostellar   &      G353.396$-$0.071 &  1.20    &   Protostellar  \\     
G008.459$-$0.222   &  0.50   &    Protostellar   &      G353.579+0.659   &  0.07    &   Protostellar  \\     
G008.684$-$0.367   &  4.48   &    Protostellar   &      G354.944$-$0.537 &  0.22    &   Protostellar  \\     
G008.706$-$0.414   &  0.51   &    Protostellar   &      G355.264$-$0.269 &  0.22    &   Protostellar  \\     
G008.736$-$0.362   &  0.77   &    Protostellar   &      G355.412+0.102   &  1.59    &   Protostellar  \\     
G009.284$-$0.147   &  0.43   &    Protostellar   &      G008.671$-$0.356 &  2.07    &   UCHII         \\     
G010.284$-$0.114   &  4.77   &    Protostellar   &      G011.109$-$0.397 &  1.38    &   UCHII         \\     
G010.404$-$0.201   &  3.26   &    Protostellar   &      G011.902$-$0.141 &  0.74    &   UCHII         \\     
G011.004$-$0.071   &  0.72   &    Protostellar   &      G012.418+0.506   &  0.64    &   UCHII         \\     
G011.082$-$0.534   &  0.38   &    Protostellar   &      G013.131$-$0.152 &  0.27    &   UCHII         \\     
G012.878$-$0.287   &  0.20   &    Protostellar   &      G331.709+0.602   &  0.79    &   UCHII         \\     
G013.178+0.059     &  1.26   &    Protostellar   &      G332.296$-$0.094 &  0.68    &   UCHII         \\     
G013.213+0.039     &  0.70   &    Protostellar   &      G335.586$-$0.291 &  0.47    &   UCHII         \\     
G013.234$-$0.071   &  0.45   &    Protostellar   &      G338.436+0.057   &  1.35    &   UCHII         \\     
G013.243$-$0.086   &  0.43   &    Protostellar   &      G340.054$-$0.244 &  1.05    &   UCHII         \\     
G013.281$-$0.321   &  0.96   &    Protostellar   &      G343.128$-$0.062 &  1.19    &   UCHII         \\     
G013.299$-$0.326   &  0.33   &    Protostellar   &      G344.424+0.046   &  1.30    &   UCHII         \\     
G013.902$-$0.516   &  0.34   &    Protostellar   &      G345.196$-$0.744 &  0.42    &   UCHII         \\     
G014.227$-$0.511   &  0.74   &    Protostellar   &      G346.076$-$0.056 &  0.98    &   UCHII         \\     
G014.777$-$0.487   &  0.53   &    Protostellar   &      G348.549$-$0.979 &  0.25    &   UCHII         \\     
G331.034$-$0.419   &  0.91   &    Protostellar   &      G351.041$-$0.336 &  0.37    &   UCHII         \\     
G331.134$-$0.484   &  1.18   &    Protostellar   &      G354.616+0.472   &  0.63    &   UCHII         \\     
G331.512$-$0.102   &  7.15   &    Protostellar   &      G005.897$-$0.444 &  1.54    &   Non           \\     
G331.571$-$0.229   &  0.09   &    Protostellar   &      G332.469$-$0.131 &  0.14    &   Non           \\     
G331.626+0.526     &  0.30   &    Protostellar   &      G332.677$-$0.614 &  2.31    &   Non           \\     
G331.709+0.582     &  1.87   &    Protostellar   &      G332.990$-$0.441 &  0.07    &   Non           \\     
G332.241$-$0.044   &  0.98   &    Protostellar   &      G337.844$-$0.376 &  0.71    &   Non           \\     
G332.276$-$0.071   &  0.38   &    Protostellar   &      G339.476+0.184   &  2.39    &   Non           \\     
G332.559$-$0.147   &  0.28   &    Protostellar   &      G352.492+0.796   &  0.15    &   Non           \\     
G332.604$-$0.167   &  0.86   &    Protostellar   &                       &          &                 \\
      \hline
     \end{tabular}
     \medskip
     NOTE. Columns are (from left to right)  Clump names, mass infall rates, Spitzer classification, Clump names, mass infall rates, Spitzer classification. The units are unit of 0.01 M$_{\odot}$ yr$^{-1}$. \\
  \end{minipage}
\end{table*}
\normalsize

In order to quantify whether the blue profile dominates in a given sample, a blue excess may defined as E: E$=$(N$_{blue}$$-$N$_{red}$)$/$N$_{total}$, where N$_{blue}$ and N$_{red}$ are the number of sources that show blue or red profiles, respectively, and N$_{total}$ is the total numbers of sample sources \citep{1997ApJ...489..719M}. The excess E values of pre-stellar, proto-stellar and UCHII clumps are 0.2857, 0.2189 and -0.1139, respectively. This result is consistent with our prediction that infall motions dominate the early stage of high-mass star formation. The corresponding probability P values (see \citet{2005A&A...442..949F}, and references therein), which characterize the blue excess arises by chance, are 0.0003, 0.00005, and 0.13, respectively.

\begin{table*}
\scriptsize
 \centering
  \begin{minipage}{125mm}
   \caption{Blue profile excess and detection rate determined from the HCO$^{+}$(1-0) line profiles for different evolutionary stages.}
    \scriptsize
     \begin{tabular}{lccccccc}
      \hline
      Classification    & N$_{blue}$ & N$_{red}$ & N$_{total}$ & N$_{infall}$ & E       & P       & D\\
      \hline
      Prestellar        & 35         & 11        & 84          & 29           & 0.2857  & 0.0003  & 0.3452\\
      Protostellar      & 85         & 41        & 201         & 78           & 0.2189  & 0.00005 & 0.3881\\
      UCHII             & 20         & 29        & 79          & 17           & -0.1139 & 0.13    & 0.2152\\
      No classification & 6          & 5         & 41          & 7            & $-$     & $-$     & $-$\\
     \hline
     \end{tabular}
    \medskip

  NOTE. Columns are (from left to right) number of blue profile clumps, number of red profile clumps, total number of clumps, number of infall candidates, the excess parameter, the probability of the distribution to arise by chance, and the detection rate for infall candidates.
  \end{minipage}
\end{table*}
\normalsize

\subsection{Physical properties of the infall candidates}
\subsubsection{Peak column density and clump mass}
Figures 7(a) and (b) show the distributions of the peak 870 $\mu$m flux densities, and their peak column densities for the infall candidates (red histogram) at different stages. The median values of the peak 870 $\mu$m flux densities for the pre-stellar, proto-stellar and UCHII clumps with infall are 1.02, 1.9 and 5.18 Jy beam$^{-1}$, respectively. The corresponding median values of H$_{2}$ column density are 4.47$\times$10$^{22}$ cm$^{-2}$, 5.50$\times$10$^{22}$ cm$^{-2}$ and 8.91$\times$10$^{22}$ cm$^{-2}$, respectively. Both distributions display an increasing trend with evolutionary stage of the clumps. The median values of the peak 870 $\mu$m flux densities for the non-infall candidates (blue histogram) at pre-stellar, proto-stellar and UCHII stages are 1.2, 2.24 and 4 Jy beam$^{-1}$, respectively. The corresponding median values of H$_{2}$ column density are 5.25$\times$10$^{22}$ cm$^{-2}$, 6.46$\times$10$^{22}$ cm$^{-2}$ and 6.92$\times$10$^{22}$ cm$^{-2}$, respectively. They also display an increase with evolutionary stage. It should be noted that the median values of the peak 870 $\mu$m flux densities and peak column densities for the infall candidates at pre-stellar and proto-stellar stages are less than the corresponding values of the non-infall candidates. Instead, the peak 870 $\mu$m flux densities and peak column densities of the infall candidates display a more obvious increasing trend with evolutionary stage. Fig. 7 (c) presents the H$_{2}$ column density distributions of all the infall candidates (red histogram), non-infall candidates (blue histogram) and the whole sample (grey histogram). The median values are 5.50$\times$10$^{22}$ cm$^{-2}$ and 6.46$\times$10$^{22}$ cm$^{-2}$ for infall and non-infall candidates, respectively. This suggests that the peak column densities of the infall candidates, in general, are smaller than those of the non-infall candidates.

The distributions of the integrated 870 $\mu$m flux densities and masses of the infall candidates (red histogram), as separated by evolutionary stage, are shown in Fig. 8 (a) and (b). The median values of the integrated 870 $\mu$m flux densities for the pre-stellar, proto-stellar and UCHII clumps with infall are 10.74, 14.59 and 34.52 Jy, respectively. The corresponding median values of their masses are 1258.9, 1659.6 and 2041.7 M$_{\odot}$, respectively. The distributions of the integrated 870 $\mu$m flux densities and masses of the non-infall candidates (blue histogram) are also shown in Fig. 8 (a) and (b). The median values of the integrated 870 $\mu$m flux densities of the pre-stellar, proto-stellar and UCHII clumps are 12.77, 15.4 and 32.5 Jy, respectively. Their corresponding median masses are 3090.3, 2238.7 and 2187.8 M$_{\odot}$, respectively.
Clearly the masses of the infall candidates increase as the clumps evolve, while those of the non-infall candidates decrease. The median mass values of the infall candidates are smaller than those of the non-infall candidates at the pre-stellar and proto-stellar stages. One reasonable explanation for these results is that a relatively large proportion of non-infall candidates at pre-stellar and proto-stellar stages have large distances, and clumps with larger distances usually have larger masses. The median values of their masses are 1659.6 M$_{\odot}$ and 2290.9 M$_{\odot}$ for all the infall and non-infall candidates, respectively (Fig. 8 (c)). The statistical parameters for each of these distributions are summarized in Table 5.

\begin{figure}
 \centering
  \includegraphics[width=0.45\textwidth]{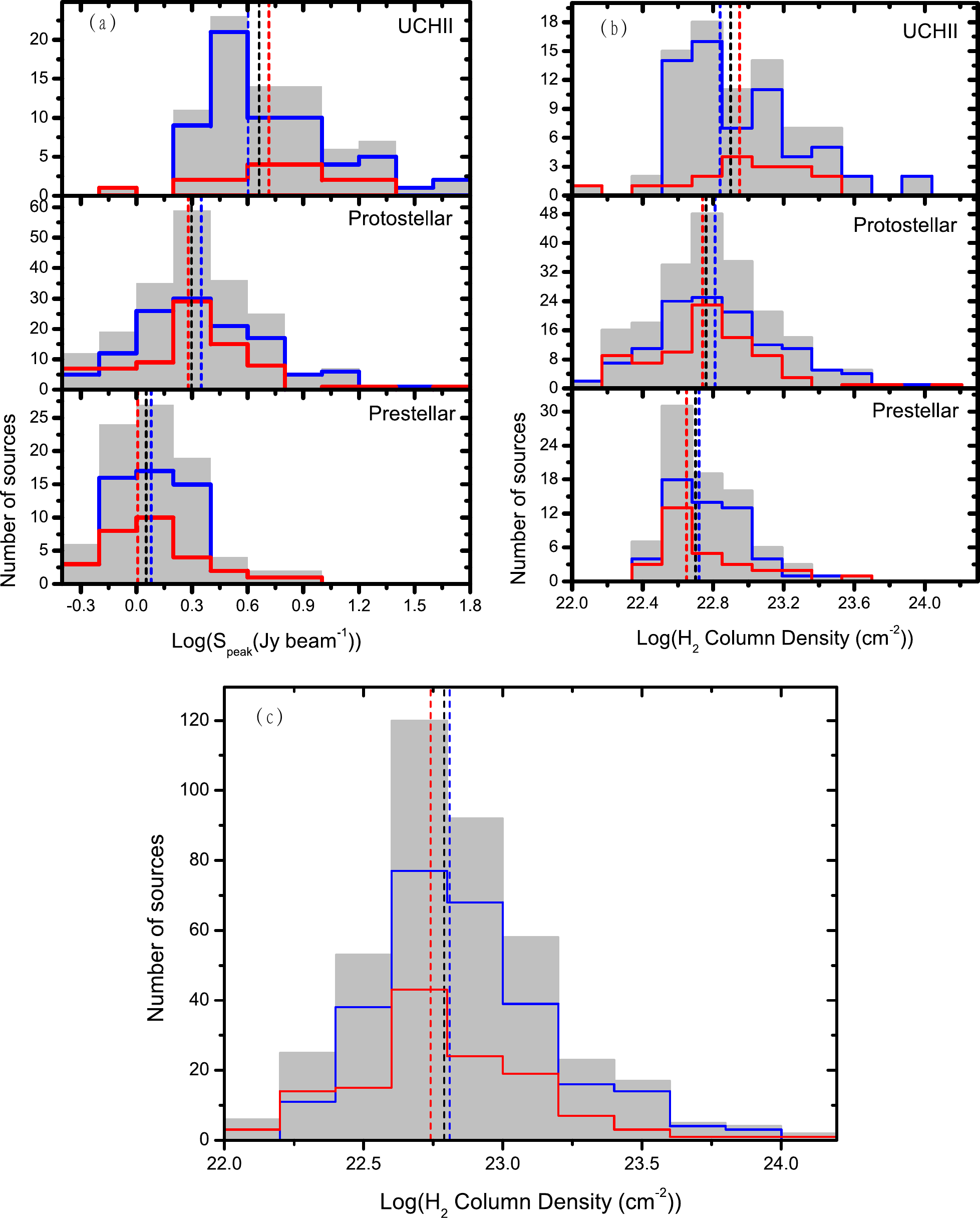}
   \caption{The 870 $\mu$m properties and H$_2$ column densities for infal and non-infall sources. (a): The 870 $\mu$m peak flux density distributions for infall candidates (red histogram), non-infall candidates (blue histogram), and all classified clumps (gray histogram) separated by evolutionary stage. The median values of each stage are indicated by dashed red, blue and black vertical lines, respectively. (b): The H$_{2}$ beam-averaged column densities derived from the 870 $\mu$m dust emission for infall candidates (red histogram), non-infall candidates (blue histogram), and all classified clumps (gray histogram) separated by evolutionary stage. The median values of each stage are indicated by dashed red, blue and black vertical lines, respectively. (c): The H$_{2}$ beam-averaged column densities distribution for the whole sample (grey histogram), infall candidates (red histogram) and non-infall candidates (blue histogram). Median values are indicated by the dashed black, red and blue vertical lines for the whole sample, infall candidates, and non-infall candidates, respectively.}
\end{figure}

\begin{figure}
 \centering
  \includegraphics[width=0.45\textwidth]{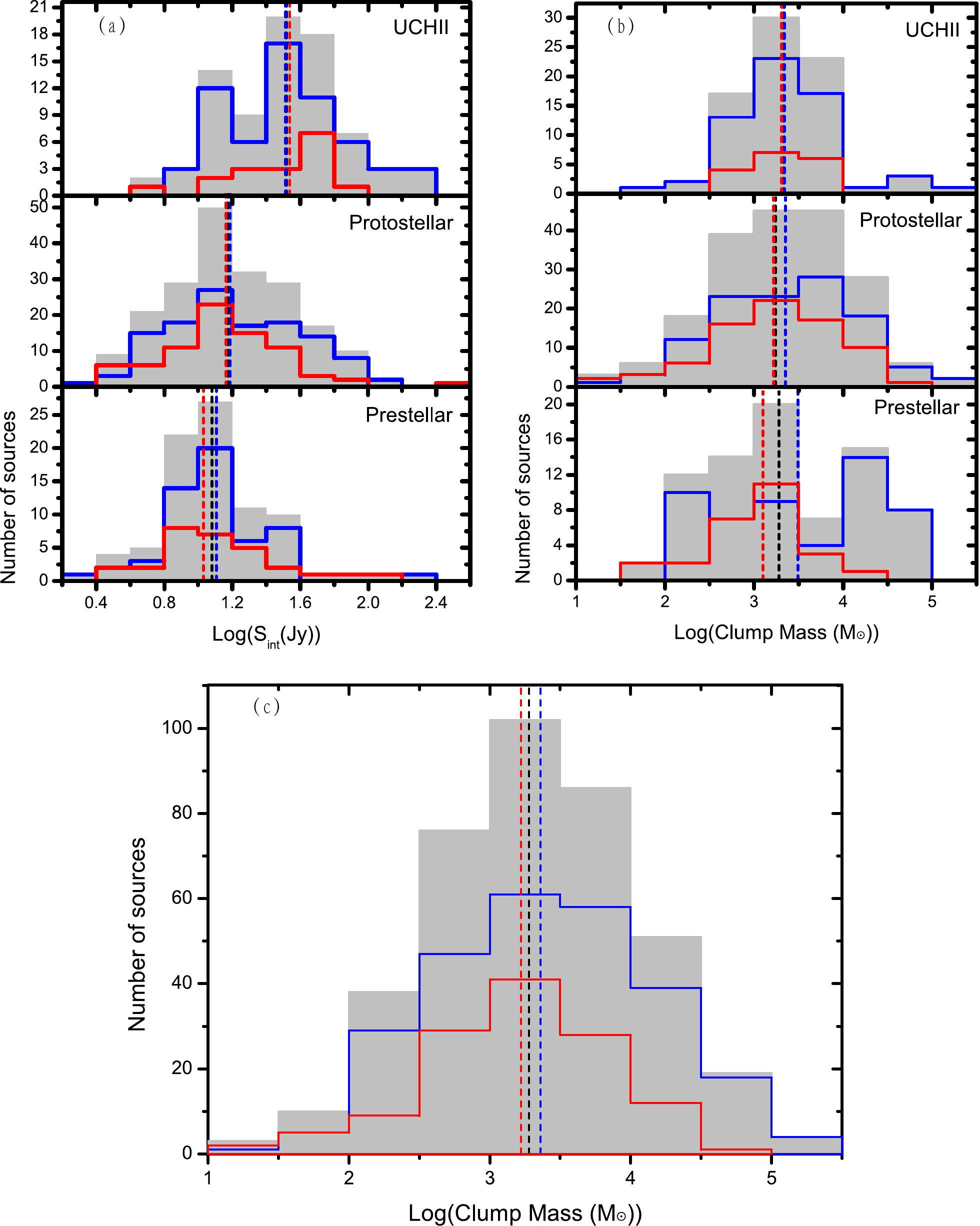}
   \caption{Same as Fig. 7, but showing 870 $\mu$m integrated flux density and clump mass.}
\end{figure}

\subsubsection{Mass infall rate}
 For the 131 infall candidates, a rough estimate of their infall rate may be determined from:    $\dot{M}$$_{inf}$ =4$\pi$R$^{2}$V$_{inf}$$\rho$ \citep[Eq. 5;][]{2010A&A...517A..66L}, where V$_{inf}$ = V$_{N_{2}H^{+}}$ $-$ V$_{HCO^{+}}$ is an estimate of the infall velocity, $\rho$=M/(4/3$\pi$R$^{3}$) is the average clump volume density, and R is the radius of the clump. Here we used R and M as derived from the dust continuum emission at 870$\mu$m. The obtained mass infall rates are listed in Table 3. Fig. 9 shows the distributions of the mass infall rates of the infall candidates separated into pre-stellar, proto-stellar and UCHII stages. Their corresponding median values are 0.0078, 0.0077 and 0.0074 M$_\odot$ yr$^{-1}$, respectively. The mass infall rates of the infall candidates display no obvious variation with evolutionary stage. The statistical parameters for each of these distributions are summarized in Table 5.

\begin{figure}
 \centering
  \includegraphics[width=0.45\textwidth]{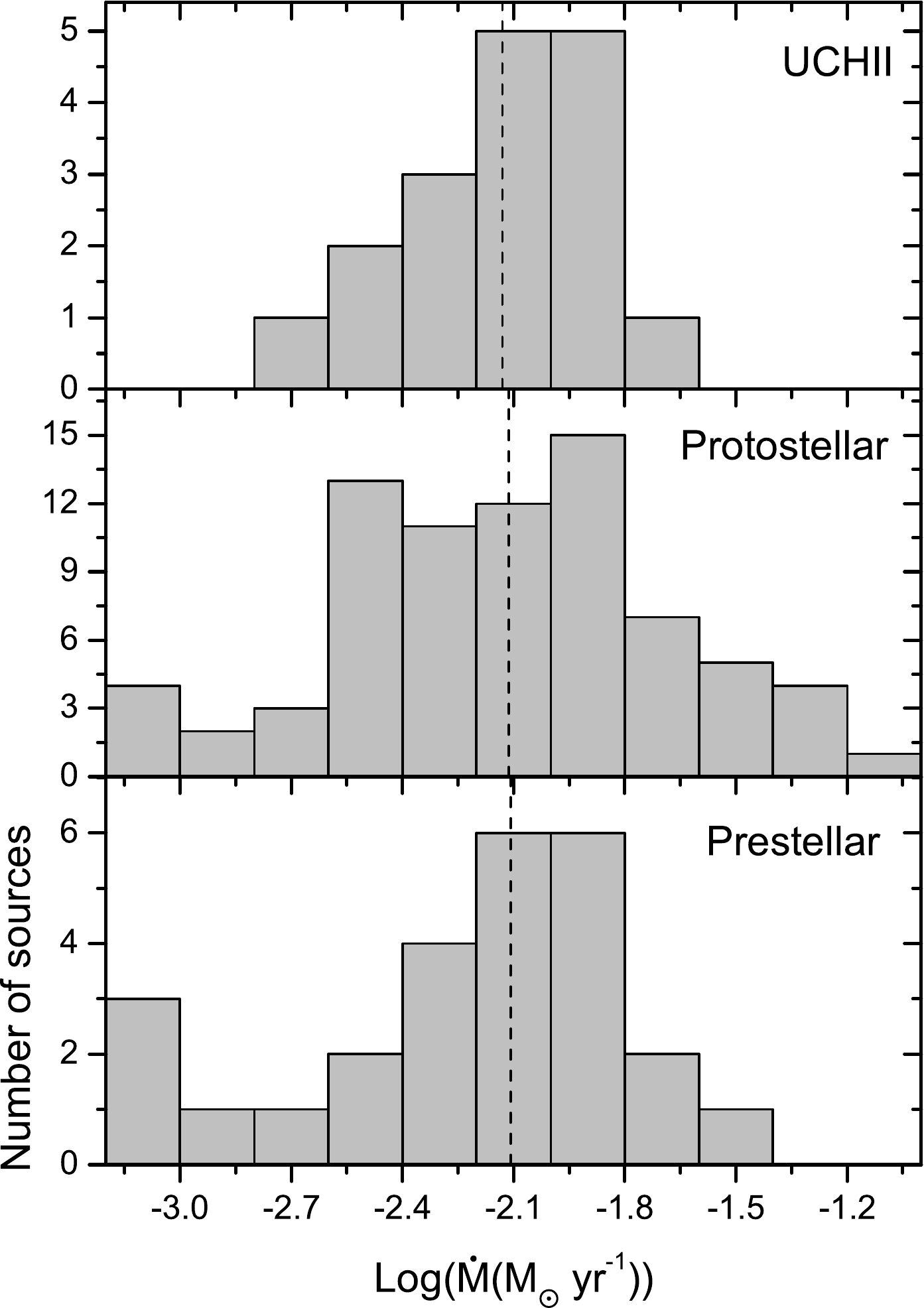}
   \caption{Mass infall rate distributions for different evolutionary stages. Median values are indicated by the dashed black vertical lines.}
\end{figure}

\begin{table*}
\scriptsize
 \centering
  \begin{minipage}{125mm}
   \caption{The mass infall rates of infall and non-infall candidates.}
    \begin{tabular}{lrrrrrrr}
      \hline
      \multicolumn{1}{c}{Property}             &  \multicolumn{1}{c}{Group}     & \multicolumn{1}{c}{Notes}     &  \multicolumn{1}{c}{Mean}    &  \multicolumn{1}{c}{Standard Deviation}   &  \multicolumn{1}{c}{Minimum}   &  \multicolumn{1}{c}{Median}  &  \multicolumn{1}{c}{Maximum} \\
      \hline
       $S_{peak}$                 &     prestellar     &   1      &       1.63         &       1.61            &          0.56    &    1.02      &    8.41       \\                         
($Jy\ beam^{-1}$)          &                    &   2      &       1.53         &       1.08            &          0.59    &    1.20      &    7.50       \\                         
                           &                    &   3      &       1.56         &       1.28            &          0.56    &    1.13      &    8.41       \\                         
                           &     protostellar   &   1      &       3.14         &       5.35            &          0.51    &    1.90      &    40.80      \\                         
                           &                    &   2      &       3.29         &       3.80            &          0.44    &    2.24      &    32.86      \\                         
                           &                    &   3      &       3.23         &       4.45            &          0.44    &    1.98      &    40.80      \\                         
                           &     UCHII          &   1      &       7.21         &       5.18            &          0.83    &    5.18      &    18.93      \\                         
                           &                    &   2      &       8.05         &       10.38           &          1.76    &    4.00      &    59.26      \\                         
                           &                    &   3      &       7.87         &       9.48            &          0.83    &    4.59      &    59.26      \\                         
Log($N(H_2)$)              &     prestellar     &   1      &       22.74        &       0.28            &          22.39   &    22.65     &    23.57      \\                         
($cm^{-2}$)                &                    &   2      &       22.77        &       0.22            &          22.42   &    22.72     &    23.52      \\                         
                           &                    &   3      &       22.76        &       0.24            &          22.39   &    22.70     &    23.57      \\                         
                           &     protostellar   &   1      &       22.77        &       0.34            &          22.17   &    22.74     &    24.07      \\                         
                           &                    &   2      &       22.82        &       0.35            &          22.11   &    22.81     &    23.98      \\                         
                           &                    &   3      &       22.80        &       0.35            &          22.11   &    22.76     &    24.07      \\                         
                           &     UCHII          &   1      &       22.97        &       0.36            &          22.15   &    22.95     &    23.51      \\                         
                           &                    &   2      &       22.96        &       0.36            &          22.48   &    22.84     &    24.01      \\                         
                           &                    &   3      &       22.96        &       0.36            &          22.15   &    22.90     &    24.01      \\                         
                           &     All sources    &   1      &       22.80        &       0.34            &          22.15   &    22.74     &    24.07      \\                         
                           &                    &   2      &       22.86        &       0.33            &          22.11   &    22.81     &    24.01      \\                         
                           &                    &   3      &       22.84        &       0.33            &          22.11   &    22.79     &    24.07      \\                         
$S_{int}$                  &     prestellar     &   1      &       19.69        &       23.92           &          2.99    &    10.74     &    107.23     \\                         
(Jy)                       &                    &   2      &       17.48        &       23.21           &          1.77    &    12.77     &    173.55     \\                         
                           &                    &   3      &       18.24        &       23.34           &          1.77    &    12.03     &    173.55     \\                         
                           &     protostellar   &   1      &       22.39        &       39.94           &          2.69    &    14.59     &    343.64     \\                         
                           &                    &   2      &       24.65        &       24.36           &          2.46    &    15.40     &    139.92     \\                         
                           &                    &   3      &       23.77        &       31.26           &          2.46    &    14.98     &    343.64     \\                         
                           &     UCHII          &   1      &       35.05        &       18.67           &          5.88    &    34.52     &    67.50      \\                         
                           &                    &   2      &       46.06        &       43.64           &          5.16    &    32.50     &    215.01     \\                         
                           &                    &   3      &       43.69        &       39.77           &          5.16    &    33.06     &    215.01     \\                         
Log($M_{clump}$)           &     prestellar     &   1      &       3.00         &       0.62            &          1.56    &    3.10      &    4.07       \\                         
($M_{\odot}$)              &                    &   2      &       3.53         &       0.87            &          2.02    &    3.49      &    4.71       \\                         
                           &                    &   3      &       3.35         &       0.83            &          1.56    &    3.28      &    4.71       \\                         
                           &     protostellar   &   1      &       3.21         &       0.72            &          1.16    &    3.22      &    4.59       \\                         
                           &                    &   2      &       3.36         &       0.78            &          1.49    &    3.35      &    5.02       \\                         
                           &                    &   3      &       3.30         &       0.76            &          1.16    &    3.24      &    5.02       \\                         
                           &     UCHII          &   1      &       3.29         &       0.34            &          2.74    &    3.31      &    3.73       \\                         
                           &                    &   2      &       3.35         &       0.58            &          1.91    &    3.34      &    5.28       \\                         
                           &                    &   3      &       3.33         &       0.54            &          1.91    &    3.33      &    5.28       \\                         
                           &     All sources    &   1      &       3.17         &       0.66            &          1.16    &    3.22      &    4.59       \\                         
                           &                    &   2      &       3.40         &       0.77            &          1.49    &    3.36      &    5.28       \\                         
                           &                    &   3      &       3.33         &       0.74            &          1.16    &    3.28      &    5.28       \\                         
$\dot{M}$                  &     prestellar     &   1      &       0.008515     &       0.006554        &          0.0007  &    0.0078    &    0.0268     \\                         
M$_{\odot}$ yr$^{-1}$      &     protostellar   &   1      &       0.012278     &       0.013092        &          0.0007  &    0.0077    &    0.0715     \\                         
                           &     UCHII          &   1      &       0.008576     &       0.004865        &          0.0025  &    0.0074    &    0.0207     \\                         
      \hline
     \end{tabular}
     \medskip
     NOTE. Column 3 notes: (1) - infall candidates, (2) - non-infall candidates, (3) - infall + non-infall candidates.\\
  \end{minipage}
\end{table*}
\normalsize

\section[]{Conclusions}
A total of 405 compact sources have been selected on the basis of the ATLASGAL and MALT90 survey data. These were then classified as pre-stellar, proto-stellar and UCHII clumps, and the optically thick lines HCO$^{+}$(1-0) and HNC(1-0), and the optically thin N$_{2}$H$^{+}$(1-0) line were used to search for infall candidates.

A total of 96.4\% of our sample sources satisfy the empirical mass-size relationship for massive star formation, and thus have the potential to form high-mass stars. Our result suggests that 0.05 g cm$^{-2}$ is a reliable lower bound for the clump surface density required for massive star formation. Among the sample, five new MPC candidates have been identified: G008.691-0.401, G333.299-0.351, G338.459+0.024, G348.183+0.482 and G348.759-0.946. These five new candidates, as well as two known MPC candidates (G345.504+0.347 and G351.774-0.537) all have large masses ($>$30.000 M$_{\odot}$), and large H$_{2}$ column densities ($>$10$^{23}$ cm$^{-2}$), except for G008.691-0.401. They all lie at far distances ($>$10 kpc), and none of these are infall candidates. The peak 870 $\mu$m flux densities and the column densities of the clumps display an increasing trend with their evolutionary stage.

A total of 131 reliable infall candidates have been identified with a detection rate towards pre-stellar, proto-stellar and UCHII clumps of 0.3452, 0.3881 and 0.2152, respectively. This supports the result that infall motions accompany the high-mass star formation process. The relatively high detection rate of the infall candidates toward the UCHII clumps indicates that many UCHII regions are still accreting matter. The roughly estimated mass infall rates of the infall candidates at pre-stellar, proto-stellar and UCHII stages are 0.0078, 0.0077 and 0.0074 M$_\odot$ yr$^{-1}$, respectively. The peak column densities and masses of the infall candidates, in general, display an increasing trend with evolutionary stage.

\section*{Acknowledgments}
This research has made use of the data products from the Millimetre Astronomy Legacy Team 90 GHz (MALT90) survey, the APEX Telescope Large Area Survey of the Galaxy (ATLASGAL) survey, which is a collaboration between the Max-Planck-Gesellschaft, the European Southern Observatory (ESO) and the Universidad de Chile, and also used NASA/IPAC Infrared Science Archive, which is operated by the Jet Propulsion Laboratory, California Institute of Technology, under contract with the National Aeronautics and Space Administration.

This work was supported by National Basic Research Program of China (973 program) No. 2012CB821802, The National Natural Science Foundation of China under grant Nos. 11373062, 11433008 and 11303081, and The Program of the Light in China's Western Region (LCRW) under grant Nos. RCPY201202 and XBBS-2014-24.
WAB acknowledges the support as a Visiting Professor of the Chinese Academy of Sciences (KJZD-EW-T01).

\end{document}